\documentclass[aps,prl,twocolumn,superscriptaddress,10pt]{revtex4-1}
\usepackage{graphicx,xcolor,longtable,physics}
\usepackage[super]{nth}
\newcommand{\Xe}{\ensuremath{^{129}\mathrm{Xe}}}
\newcommand{\He}{\ensuremath{^{3}\mathrm{He}}}
\begin{document}

\title{New Limit on the Permanent Electric Dipole Moment of $^{129}$Xe using $^{3}$He Comagnetometry and SQUID Detection}

\author{N.~Sachdeva}
\email{sachd@umich.edu}
\affiliation{Department of Physics, University of Michigan, Ann Arbor, Michigan 48109, USA}
\author{I.~Fan}
\affiliation{Physikalisch-Technische Bundesanstalt (PTB) Berlin, 10587 Berlin, Germany}
\author{E.~Babcock}
\affiliation{J\"ulich Center for Neutron Science, 85748 Garching, Germany}
\author{M.~Burghoff}
\affiliation{Physikalisch-Technische Bundesanstalt (PTB) Berlin, 10587 Berlin, Germany}
\author{T.~E.~Chupp}
\affiliation{Department of Physics, University of Michigan, Ann Arbor, Michigan 48109, USA}
\author{S.~Degenkolb}
\affiliation{Department of Physics, University of Michigan, Ann Arbor, Michigan 48109, USA}
\affiliation{Institut Laue-Langevin, 38042 Grenoble, France}
\author{P.~Fierlinger}
\affiliation{Excellence Cluster Universe and Technische Universit{\"a}t M{\"u}nchen, 85748 Garching, Germany}
\author{S.~Haude}
\affiliation{Physikalisch-Technische Bundesanstalt (PTB) Berlin, 10587 Berlin, Germany}
\author{E.~Kraegeloh}
\affiliation{Excellence Cluster Universe and Technische Universit{\"a}t M{\"u}nchen, 85748 Garching, Germany}
\affiliation{Department of Physics, University of Michigan, Ann Arbor, Michigan 48109, USA}
\author{W.~Kilian}
\affiliation{Physikalisch-Technische Bundesanstalt (PTB) Berlin, 10587 Berlin, Germany}
\author{S.~Knappe-Gr{\"u}neberg}
\affiliation{Physikalisch-Technische Bundesanstalt (PTB) Berlin, 10587 Berlin, Germany}
\author{F.~Kuchler}
\affiliation{Excellence Cluster Universe and Technische Universit{\"a}t M{\"u}nchen, 85748 Garching, Germany}
\affiliation{TRIUMF, Vancouver, British Columbia V6T 2A3, Canada}
\author{T.~Liu}
\affiliation{Physikalisch-Technische Bundesanstalt (PTB) Berlin, 10587 Berlin, Germany}
\author{M.~Marino}
\affiliation{Excellence Cluster Universe and Technische Universit{\"a}t M{\"u}nchen, 85748 Garching, Germany}
\author{J.~Meinel}
\affiliation{Excellence Cluster Universe and Technische Universit{\"a}t M{\"u}nchen, 85748 Garching, Germany}
\author{K.~Rolfs}
\affiliation{Physikalisch-Technische Bundesanstalt (PTB) Berlin, 10587 Berlin, Germany}
\author{Z.~Salhi}
\affiliation{J\"ulich Center for Neutron Science, 85748 Garching, Germany}
\author{A.~Schnabel}
\affiliation{Physikalisch-Technische Bundesanstalt (PTB) Berlin, 10587 Berlin, Germany}
\author{J.~T.~Singh}
\affiliation{National Superconducting Cyclotron Laboratory and Department of Physics \& Astronomy, Michigan State University, East Lansing, Michigan 48824, USA}
\author{S.~Stuiber}
\affiliation{Excellence Cluster Universe and Technische Universit{\"a}t M{\"u}nchen, 85748 Garching, Germany}
\author{W.~A.~Terrano}
\affiliation{Excellence Cluster Universe and Technische Universit{\"a}t M{\"u}nchen, 85748 Garching, Germany}
\author{L.~Trahms}
\affiliation{Physikalisch-Technische Bundesanstalt (PTB) Berlin, 10587 Berlin, Germany}
\author{J.~Voigt}
\affiliation{Physikalisch-Technische Bundesanstalt (PTB) Berlin, 10587 Berlin, Germany}

\date{\today}

\begin{abstract}
We report results of a new technique to measure the electric dipole moment of $^{129}$Xe with $^3$He comagnetometry. Both species are polarized using spin-exchange optical pumping, transferred to a measurement cell, and transported into a magnetically shielded room, where SQUID magnetometers detect free precession in applied electric and magnetic fields. The result from a one week measurement campaign in 2017 and a 2.5 week campaign in 2018, combined with detailed study of systematic effects, is $d_A(\Xe) = (1.4 \pm 6.6_\mathrm{stat} \pm 2.0_\mathrm{syst})\times10^{-28}~e\,\mathrm{cm}$.  This corresponds to an upper limit of $|d_A(\Xe)| < 1.4 \times 10^{-27} ~e\,\mathrm{cm}~(95\%~\mathrm{CL})$, a factor of five more sensitive than the limit set in 2001.
\end{abstract}
%\keywords{}
\maketitle %consistency, due to, in all cases, checks
Searches for permanent electric dipole moments (EDMs) are a powerful way to investigate beyond-standard-model (BSM) physics. An EDM is a charge asymmetry along the total angular momentum axis of a particle or system and is odd under both parity reversal (P) and time reversal (T). Assuming CPT conservation (C is charge conjugation), an EDM is a direct signal of CP violation (CPV), a condition required to generate the observed baryon asymmetry of the universe~\cite{Sakharov1967}. The Standard Model incorporates CPV through the phase in the CKM matrix and the QCD parameter $\bar{\theta}$. However, the Standard Model alone is insufficient to explain the size of the baryon asymmetry~\cite{Dine2003}. BSM scenarios that generate the observed baryon asymmetry~\cite{Morrissey2012} generally also provide for EDMs larger than the SM estimate, which for $\Xe$ is $|d_A(\Xe) ^\mathrm{SM}|\approx 5\times 10^{-35}~e\,$cm~\cite{Chupp2017}. 

EDM measurements have provided constraints on how BSM CPV can enter low-energy physics \cite{Chupp2017}. Diamagnetic systems such as $\Xe$ and $^{199}$Hg are particularly sensitive to CPV nucleon-nucleon interactions that induce a nuclear Schiff moment and CPV semileptonic couplings~\cite{Chupp2015}. While the most precise atomic EDM measurement is from $^{199}$Hg~\cite{Graner2016}, there are theoretical challenges to constraining hadronic CPV parameters from $^{199}$Hg alone, and improved sensitivity to the $\Xe$ EDM would tighten these constraints~\cite{Chupp2015, Yamanaka2017}. Additionally, recent work has shown that contributions from light-axion-induced CPV are significantly stronger for $\Xe$ than for $^{199}$Hg~\cite{DzubaFlambaum2018}. $\Xe$ also may be used as a comagnetometer in future neutron EDM experiments~\cite{Degenkolb2012,Masuda2012}.

The first $\Xe$ EDM measurement by Vold \textit{et al.} monitored $\Xe$ Larmor precession frequency as a function of applied electric field~\cite{Vold1984}. Rosenberry \textit{et al.}~\cite{Rosenberry2001} used a two-species maser with a $\He$ comagnetometer. A number of $^{129}$Xe EDM efforts to improve on this limit have followed, including an active maser technique~\cite{Inoue2016}, and an experiment with polarized liquid xenon~\cite{Ledbetter2012}. Recently the result of an experiment using $^{3}$He and SQUID detection, but with a different approach to EDM extraction and systematic effects, was reported~\cite{Allmendinger2019}. The early developments of our approach are described in Ref.~\cite{Kuchler2016}.

For a system with total angular momentum $\vec F$, EDM $d\vec F/F$, and magnetic moment $\mu\vec F/F$, the Hamiltonian is $H = -(\mu \vec F\cdot\vec B + d\vec F\cdot\vec E)/F$. This results in an energy splitting dependent on $\vec E\cdot \hat B$ and a corresponding frequency shift $\omega_d =\pm d\,|E|/(\hbar F)$ between states with $|\Delta m_F|=1$. Changes of $\vec B$ due to drifts and extraneous magnetic fields lead to frequency shifts that are mitigated by comagnetometry---simultaneous measurement with a colocated species. The $^{129}$Xe-$^3$He comagnetometer system is favorable because both can be simultaneously polarized by spin-exchange optical pumping (SEOP) \cite{Chupp1988}, have long spin relaxation times enabling precision frequency measurements, and $\He$, with $27\times$ lower nuclear charge $Z$, is much less sensitive to CP violation~\cite{FlambaumGinges}. 

\begin{figure}[!t]
\centering
\vskip -0.14 truein
\includegraphics[width=\columnwidth]{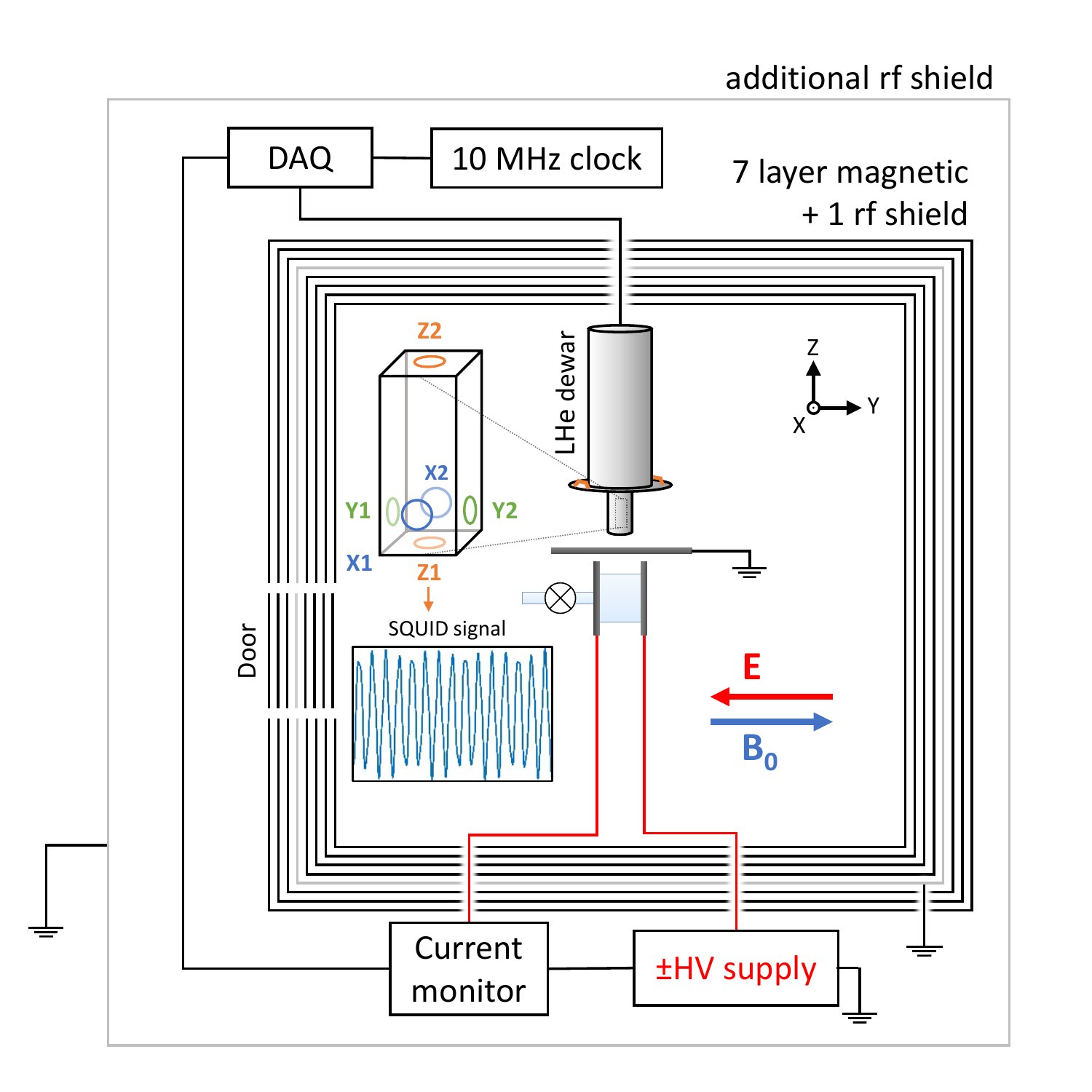}
\vskip -0.14 true in
\caption{\label{apparatus} (color online) Schematic of the HeXeEDM apparatus at PTB. The electric field $\vec E$ indicated corresponded to +HV and the magnetic field is shown along $+\hat y$ for 2017. In 2018, the electric and magnetic fields were along $\hat{x}$. The inset shows a typical raw SQUID signal for 1/2 second of data; the frequencies were 30.8~Hz (35.2~Hz) for $\Xe$ and 84.8~Hz (97.0~Hz) for $\He$ in 2017 (2018). Not to scale.}
\vskip -0.25 true in
\end{figure}

We present the combined results of two HeXeEDM campaigns in 2017 and 2018 at the BMSR-2 (Berlin Magnetically Shielded Room) facility at Physikalisch-Technische Bundesanstalt (PTB) Berlin. The layout of the experiment is shown in Fig.~\ref{apparatus}. Free precession of $\Xe$ and $\He$ was measured with low-noise superconducting quantum interference devices (SQUIDs). The BMSR-2 provided a passive shielding factor of more than $10^8$ above 6 Hz~\cite{Bork2002}. A 1.6~m diameter set of Helmholtz coils generated the static magnetic field ($B_0$) of 2.6--3.0~$\mathrm{\mu T}$ along the $y$-axis in 2017 and $x$-axis in 2018. In a separate setup similar to that described in Ref.~\cite{Korchak2013}, for the 2017 (2018) campaign, the gas mixture of 18\% (15\%) isotopically enriched xenon ($90\%$~$\Xe$), $73\% (75\%)~\He$, and $9\% (15\%)~\mathrm{N_2}$ was polarized by SEOP in a refillable optical pumping cell (OPC). Simultaneous polarization of $\Xe$-$\He$ mixtures compromise both polarizations because the optimum conditions are very different for the two species. Typically, we achieved 5--15\% polarization for $\Xe$ and 0.1--0.2\% (2017) or 0.5--1.4\% (2018) polarization for $\He$ depending on the total pressure in the OPC. Data were taken with three EDM cells with 30~mm diameter, 2~mm thickness, p-type (Boron) doped 1-10 $\Omega$\,cm silicon electrodes diffusion bonded to borosilicate glass cylinders~\cite{PatrickPistelPC}. One cell (PP1) had a length of 18.5~mm and an inner diameter of 20.5~mm; PP2 and PP3 both had a length of 21.8~mm and an inner diameter of 20.4~mm. PP1 and PP2 were used in 2017; all three cells were used in 2018. %Both cells had slightly larger length-to-diameter ratios~\cite{Limes2018} than would be optimum to suppress comagnetometer drifts, which are discussed below. 
Before each filling, the EDM cell was degaussed using a commercial bulk degausser~\cite{degausser}. The polarized gas was expanded from the OPC into an evacuated EDM cell. Each time the OPC was refilled, the polarized gas was used for two EDM cell fillings: the first had higher pressure ($\sim$1~bar) and the second had lower pressure ($\sim$0.5~bar). Toward the end of the 2018 campaign, we shifted to using only higher pressures in a scheme that prioritized $\Xe$ polarization, resulting in improved SNR and a reduction of the comagnetometer drift discussed below. After the EDM cell was filled, it was transported to the magnetically-shielded room in a battery-powered $400~\mathrm{\mu T}$ shielded solenoid and positioned under the SQUID dewar. 
%After the BMSR-2 door was closed, the magnetic field was allowed to stabilize for about five minutes. 
For the 2017 campaign, we applied a time-dependent magnetic field along the $x$-axis with resonant frequency components and amplitudes tuned to effect a $\pi/2$ pulse for both species. For 2018, the magnetic field was diabatically switched within 0.5~ms from $\hat y$ to $\hat x$. For both campaigns, data were acquired from the $Z_1$-SQUID, which was located 50~mm (2017) and 36~mm (2018) above the center of the EDM cell. A grounded silicon wafer was placed between the EDM cell and dewar as indicated in Fig.~\ref{apparatus} to protect the SQUIDs from HV discharges.

The data-acquisition sample rate of 915.5245~Hz was derived from the 10~MHz output of an external clock~\cite{quartzclock}. The initial amplitudes of the precession signals were about 30~pT and 5~pT for $^{129}$Xe and $^3$He, respectively, in 2017, and 20--70~pT and 17--50~pT in 2018. The noise measured by the SQUID system was $6~\mathrm{fT/\sqrt{Hz}}$. The free precession decay time $T_2^*$ did not differ significantly between the two campaigns and was in the range of 3000--10000~s for $\Xe$ and 4000--10000~s for $\He$. The precession was measured typically for about 15,000~s in 2017 and 25,000--45,000 seconds in 2018, which we define as a {\it run}. During each run, a pattern of HV polarity changes modulated the EDM signal. A pattern with changes in equal length intervals defines a {\it subrun}. In 2018, there were 1--4 subruns within a run with different segment lengths; in 2017, there was only one subrun per run. 

HV of $\pm\,$6--9~kV was applied to one electrode with the other electrode connected via the current monitor to ground potential. In 2017, the average electric fields were 3.2~kV/cm and 2.7~kV/cm across cells PP1 and PP2, respectively. In 2018, the electric fields ranged from 2.7~kV/cm to 4.8~kV/cm. The voltage was chosen to be below the observed breakdown voltage.

During each subrun, the HV polarity was positive ($+$), negative ($-$), or zero for equal intervals called segments. Segments with zero HV were inserted at the beginning and end of each set of 16 segments within a subrun~\cite{Sachdeva2019}. The rate of change of HV between segments (HV ramp) was set to either 1 or 2 kV/s in 2017 and 0.5 or 1~kV/s in 2018. Segment lengths of 400 or 800 seconds for 2017 and between 100 to 600 seconds for 2018 were chosen based on the Allan deviation minimum from studies before taking EDM data. During analysis, an $F$-test was used to check for comagnetometer drift within segments. In 2017, five segments out of a total of 539 segments were shortened accordingly due to comagnetometer drift and an additional eight were shortened because of HV or SQUID problems. For 2018, four runs were removed: three due to HV spark and SQUID irregularities and one due to a large magnetic field shift halfway through the run.

The raw time-domain SQUID data were processed by first removing the DC offset and baseline drift with a high-pass filter. Filtered data were divided into non-overlapping blocks of length $\tau = 20~$seconds, short enough that amplitude decay and frequency drift were negligible. Data for each block were fit using a separable nonlinear least-squares method \cite{Golub2003} to a six-parameter model to determine the amplitude, frequency, and phase $\Phi^m_\mathrm{Xe/He}$ for block $m$ for each species (see supplement \cite{HeXeSupplement}). An independent analysis was performed using an alternative approach, which did not use the high-pass filter but added an offset and linear drift term to the fit function as described in~\cite{Gemmel2010}, and produced consistent results. 

Magnetic field drifts were compensated by the comagnetometer corrected phases $\Phi^m_\mathrm{co} = \Phi^m_{\mathrm{Xe}} - R\Phi^m_{\mathrm{He}}$, where $R=1/2.7540816$ is the nominal ratio of the shielded gyromagnetic ratios of $\Xe$ and $\He$~\cite{Fan2016}. For each HV segment, the comagnetometer frequency $\omega_\mathrm{co}$ and uncertainty were determined from the slope of a linear fit to $\Phi^m_{\mathrm{co}}$ as a function of time. The frequency uncertainties were consistent with the minimum expected uncertainties based on the signal amplitude, noise, and segment duration~\cite{Chupp1994,Chibane1995,Sachdeva2019}. Segment frequencies were blinded by adding or subtracting, depending on the sign of $\vec E\cdot\hat B$, an unknown offset derived from a previously computer-generated pseudorandom number such that $|\omega_\mathrm{blind}^{2017}|/(2\pi)\le$ 50~nHz or $|d_\mathrm{blind}^{2018}|\le$ $5\times 10^{-27}~e\,\mathrm{cm}$. The blinding offsets were saved separately from the data in a binary format. After all cuts and systematic corrections were determined, the blinding offset was set to zero to produce the set of HV segment frequencies for the final unblinded EDM analysis.

The EDM frequency was determined from an average of four consecutive segment frequencies with HV $(+--+)$ or $(-++-)$ to compensate for linear drifts of the comagnetometer frequencies, typically a few $\mu$Hz over the course of a run. The EDM for each subrun was determined from the weighted average of the 4-segment EDM frequencies within the subrun.

Systematic effects include the uncertainties of experimental parameters as well as false-EDM signals that may arise from the nonideal response of the comagnetometer. The comagnetometer frequency $\omega_\mathrm{co}$ can be described by the following four dominant terms plus the EDM contribution $\omega_d\equiv\omega_{d_\mathrm{Xe}}-R\omega_{d_\mathrm{He}}$:
\begin{eqnarray}\label{comagmodel}
\omega_\mathrm{co}&\approx\ &\omega_{d}-\gamma^\prime_\mathrm{He}\Delta R B + \left (1-R \right)\vec\Omega\cdot\hat B \nonumber\\
&+& \gamma^\prime_\mathrm{Xe}\left (\Delta B^\mathrm{dif}_\mathrm{Xe}-\Delta B^\mathrm{dif}_\mathrm{He}\right )
+  \left (\omega^{sd}_\mathrm{Xe}-R\omega^{sd}_\mathrm{He} \right ).
\end{eqnarray}
Here, $\gamma^\prime_\mathrm{He/Xe}$ are the shielded gyromagnetic ratios; $\Delta R$ is a correction to $R$ that changed from run to run due mostly to pressure dependence of the chemical shifts; $\vec B$ is the average magnetic field within the cell with contributions from the applied magnetic field $\vec B_0$, the ambient magnetic field of the room, and any nearby magnetized materials; $\vec\Omega$ is the angular frequency of the Earth's rotation; and $\Delta B^\mathrm{dif}_\mathrm{Xe/He}$ represents the difference of the volume averaged magnetic field and the field averaged by the atoms of each species as they diffuse through the cell. In the presence of second- and higher-order gradients, this average is different for the two species~\cite{Sheng2014}. 

The $\nth{2}$ through $\nth{4}$ terms in Eq.~\ref{comagmodel} indicate the residual sensitivity of $\omega_\mathrm{co}$ to the magnitude, direction, and gradients of the magnetic field, and any correlation of these with the HV may cause a false-EDM signal. Such correlations are expected from possible leakage currents, magnetization induced by charging currents that flow when the HV is changed, and motion of the measurement cell due to electrostatic forces. Our approach to estimating false-EDM signals is based on auxiliary measurements of the dependence of $\omega_\mathrm{co}$ on amplified leakage and charging currents, gradients, and cell motion, which are scaled to the HV correlations of these parameters monitored during the experiment. The last term in Eq.~\ref{comagmodel} reflects time-dependent, species-dependent shifts, predominantly due to effects of residual longitudinal magnetization that dominate the comagnetometer drift~\cite{Terrano2018,Limes2018}. Eq.~\ref{comagmodel} does not include $\vec E \times \vec v$ effects, which are negligible. 

\begin{table}[!t] %add [H] placement to break table across pages
%\begin{ruledtabular}
%\centering
\begin{tabular}{l r r}
  & \textbf{2017($e\,\mathrm{cm}$)} & \textbf{2018 ($e\,\mathrm{cm}$)} \\ \hline
 \textbf {EDM} & $7.2\times 10^{-28}$ & $0.9\times 10^{-28}$\\
 \textbf {Statistical error} & $23.5\times 10^{-28}$ & $6.8\times 10^{-28}$\\\hline
  \textbf{Systematic Source}  &  & \\ 
Leakage current &  $1.2 \times 10^{-28}$ & $4.5\times 10^{-31}$ \\ 
Charging currents & $1.7\times 10^{-29}$ & $1.2\times 10^{-29}$ \\ 
Cell motion (rotation)&$4.2\times10^{-29}$ & $4.0\times10^{-29}$\\ 
Cell motion (translation)&$2.6\times10^{-28}$ &$1.9\times10^{-28}$ \\ 
Comagnetometer drift & $2.6\times10^{-28}$ & $4.0\times10^{-29}$\\
$|\vec E|^2$ effects &  $1.2\times10^{-29}$& $2.2\times10^{-30}$\\ 
$|\vec E|$ uncertainty & $2.6\times10^{-29}$ & $9.4\times10^{-30}$\\
Geometric phase & $\le 2\times10^{-31}$ & $\le 2\times10^{-31}$ \\ 
 \hline
Total Systematic Error & $3.9\times10^{-28}$ & $2.0\times10^{-28}$\\
\end{tabular}
%\end{ruledtabular}
%\vskip -0.1 truein                     
\caption{\label{systematicstable} Summary of EDM results and systematic effects discussed in the text.}
\vskip -0.1 truein
\end{table}

%\begin{table}[t] %add [H] placement to break table across pages
%\begin{ruledtabular}
%\centering
%\begin{tabular}{l c}
 %\textbf{Source}  & \textbf{Sys. Error ($e\,\mathrm{cm}$)} \\ \hline
%Leakage current &  $1.2 \times 10^{-28}$ \\ 
%Charging currents & $1.7\times 10^{-29}$\\ 
%$\vec E$-correlated cell motion (rotation)&$4.2\times10^{-29}$ \\ 
%$\vec E$-correlated cell motion (translation)&$2.6\times10^{-28}$ \\ 
%Comagnetometer drift & $6.6\times10^{-28}$ \\
%$|\vec E|^2$ effects &  $1.2\times10^{-29}$\\ 
%$|\vec E|$ uncertainty & $2.6\times10^{-29}$  \\
%Geometric phase & $\le 2\times10^{-31}$ \\ 
 %\hline
%Total & $7.2\times10^{-28}$ 
%\end{tabular}
%\end{ruledtabular}
%\vskip -0.1 truein                     
%\caption{\label{systematicstable} Summary of false EDM and other systematic effects discussed in the text.}
%\vskip -0.1 truein
%\end{table}
Systematic effects, including false EDM contributions and their uncertainties for both campaigns, are listed in Table~\ref{systematicstable}. During each campaign, an auxiliary measurement of the comagnetometer response to a leakage current was simulated by a single turn of wire wrapped around the cell and scaled by the observed maximum leakage current of 97~pA in 2017 and 73~pA in 2018. Since the leakage current followed an unknown path that could increase or decrease $B$, we consider this an upper limit on the magnitude of a false EDM. During each HV ramp, the charging current might have induced magnetization of materials in or near the cell, correlated with the change of HV. The comagnetometer response to charging currents of $\pm10~\mu$A and $\pm 20~\mu$A was measured and scaled by the maximum charging current observed for the EDM data. 

The electric force between the cell electrodes and the grounded safety electrode might have caused cell movement when the electric field was changed, affecting the magnetic fields and gradients across the cell. The effect of cell rotation on the comagnetometer frequency was measured by rotating the cell $\pm~5^\circ$ around the $z$-axis. HV-correlated cell rotation was investigated by measuring the motion of a laser beam spot reflected from the cell electrode with a lever arm of $1.5~$m and estimated to be less than 33~$\mu$rad. HV-correlated translation of the cell in a nonuniform magnetic field might produce a false EDM because of the change of $B$ in the cell ($\nth{2}$ term in Eq.~\ref{comagmodel}) or through a change of the higher-order gradients ($\nth{4}$ term in Eq.~\ref{comagmodel}). %The $B$ dependence of $\omega_\mathrm{co}$ was estimated from the change in chemical shift for different cell pressures. %The HV-correlated amplitude of the spin-precession signals was used to estimate a limit of 30~$\mu$m on cell translation with respect to the SQUIDs. %Combined with the linear magnetic field gradient based on $T_2^*$ for the two species~\cite{McGregor1990}, we determined the upper limit on the false EDM due to translation in a linear gradient. 
The $\nth{4}$ term in Eq.~\ref{comagmodel} is dominant and was isolated with an auxiliary measurement of $\frac{\partial\omega_\mathrm{co}}{\partial\omega_\mathrm{He}}$ for a loop mounted on a cell electrode combined with $\delta \omega_\mathrm{He}$. This provided an upper limit on any HV correlated effect, including cell translation, due to a source of magnetic field gradient outside the cell, provided the size of the source was smaller than its distance from the cell \cite{HeXeSupplement}. 

Uncompensated drift of $\omega_\mathrm{co}$ would appear as a false EDM due to the frequency shift between segments with opposite $\vec E\cdot \hat B$. The time dependence of the comagnetometer frequency drifts for all subruns could be accurately parametrized by polynomials of \nth{1} through \nth{5} order depending on the size of the drift and the signal-to-noise ratio. Offsets and linear drifts were compensated by the four-segment HV reversal pattern, while drifts characterized by \nth{2} and \nth{3} order time dependence would be removed by the eight and 16-segment HV patterns, respectively. Because the linear time dependence is dominant, we have chosen to extract the EDM using four-segment measurements ($+--+$ or $-++-$) and to apply a correction for quadratic and higher order time dependence. The correction was estimated from the weighted polynomial coefficients of the fits to the comagnetometer frequency drift for each subrun. The highest polynomial order needed to accurately parametrize the drift for each subrun was determined by applying an $F$-test. %We studied the dependence of the correction on the threshold $F_\mathrm{min}$ for $\int_{F_\mathrm{min}}^\infty P(F)dF$ and found corrections smaller than the uncertainty due to the fit parameters in all cases. 
A threshold of $F_\mathrm{min} = 0.6$ was chosen for both data sets. The uncertainty on this correction is a statistical error based on the polynomial fits to the segment frequencies for each run, but is compiled as a systematic error in Table~\ref{systematicstable} to emphasize that it may give rise to a false EDM. For the 2018 analysis, we applied the comagnetometer drift correction to the EDM for each subrun and included correlations between coefficients. Applying this method to the 2017 data resulted in a shift of the central value reported in~\cite{Sachdeva2019,Sachdeva2019arXiv} by approximately the estimated systematic uncertainty. % A more detailed description of comagnetometer drift and correction is presented in~\cite{HeXeSupplement}.

$|\vec E|^2$ effects included any shift that depended on the magnitude of the applied electric field, for example, chemical shifts or HV-induced noise detected by the SQUID. Segments with $E=0$ and the different $E$ enabled studies of the correlation of comagnetometer frequency with $\abs{E}$ and $|\vec E|^2$. The modeling of the average electric field in the cell in the presence of the protection electrode contributed an uncertainty of $0.1d_A$. The combination of $\vec E\times \vec v$ effects coupled with magnetic field gradients could produce a false EDM, often referred to as a geometric phase. In gases at the densities used for these experiments, the time between collisions is small compared to the spin-precession period, which mitigates the coherent build up of phase linear in the electric field. The formalism of Ref.~\cite{Pignol2015} was used to estimate an upper limit. 

\begin{figure}[!t]
\centering
\includegraphics[width=\columnwidth]{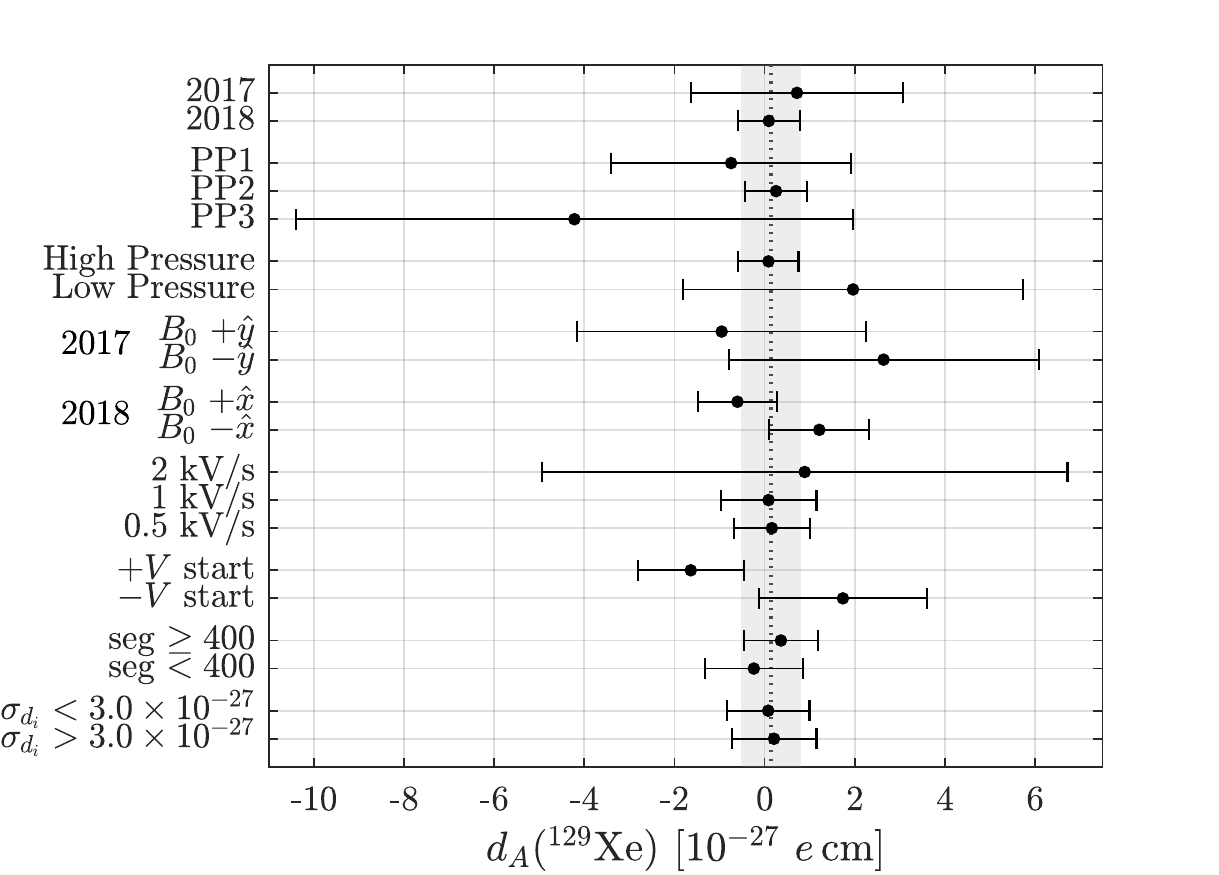}
\vskip -0.05 truein
\caption{\label{correlationsfig} Comparison of EDM measurements for both the 2017 and 2018 campaigns grouped by cell, cell pressure, $\hat B_0$ direction, HV ramp rate, HV start polarity, HV segment length, and EDM uncertainty $\sigma_{d_i}<3.00\times 10^{-27}~e\,\mathrm{cm}$ (5 subruns) or $\sigma_{d_i}>3.00\times 10^{-27}~e\,\mathrm{cm}$ (75 subruns). The shaded area shows the result given in Eq.~\ref{result}. }
\vskip -0.195 truein
\end{figure}

\begin{figure}[tb] %8.6cm
\centering
\vskip 0.05 truein
\includegraphics[width=\columnwidth]{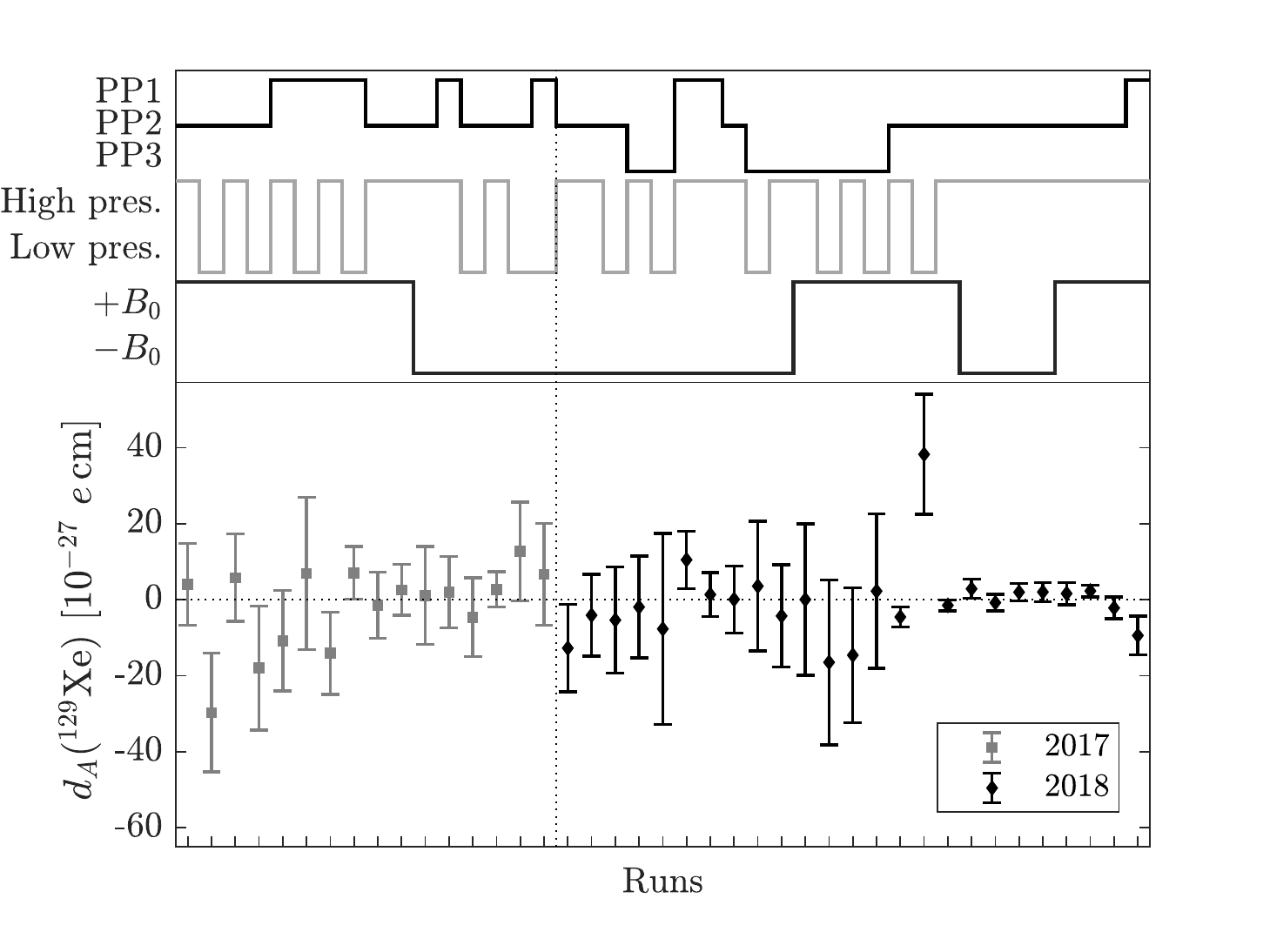}
\vskip -0.2 truein
\caption{\label{dataquality}  All drift-corrected EDM measurements by run indicating the cell used, cell pressure, and the magnetic field direction. During the 2018 run, an adjusted polarization routine resulted in a reduction of the comagnetometer drift allowing for longer segments and increased SNR. Therefore, runs from the last week of data collection had improved statistical sensitivity.}
\vskip -0.16 truein
\end{figure}

In 2017 and 2018, respectively, there were a total of 16 runs/subruns and 25 runs (64 subruns) measured under different conditions including measurement cell, gas pressure, $\vec B_0$ direction, HV ramp rate, starting HV polarity, and HV segment length. Fig.~\ref{correlationsfig} shows a comparison of sorting all EDM measurements into groups based on these variables, including EDM uncertainty $\sigma_{d_i}$, and Fig.~\ref{dataquality} shows the EDM measurements per run that had different cells, cell pressures, and orientations of $\vec B_0$ for 2017 and 2018. We also investigated correlations between the extracted EDM and other parameters including $T_2^*$ and comagnetometer drift rate~\cite{HeXeSupplement}.

%The consistency of EDM measurements over the variety of conditions and cuts illustrated in Fig.~\ref{correlationsfig} justified taking the weighted average of the EDM measurements, providing the
The comagnetometer-drift corrected results for 2017 and 2018 were confirmed with two independent analyses and are presented in Table~\ref{systematicstable}. The combined result is 
\begin{equation}
d_A(\Xe) = (1.4 \pm 6.6\ (\mathrm{stat}))\times 10^{-28}~e\,\mathrm{cm}.
\label{result}
\end{equation}
The statistical error is the uncertainty of the weighted average of the uncorrected measurements, and $\chi^2=68$ for 79 D.F. Combined with the systematic error from Table~\ref{systematicstable}, we find $|d_A(\Xe)| \le 1.4\times 10^{-27}~e\,\mathrm{cm}$ (95\% CL). This is a factor of five improvement in sensitivity over the previous limit of $|d_A(\Xe)| \le 6.6 \times 10^{-27}~e\,\mathrm{cm}$ (95\% CL)~\cite{Rosenberry2001}. Bootstrapping \cite{Efron1982} the unblinded 2017 and 2018 subrun data to estimate the error on the mean resulted in an estimate of $7.4 \times 10^{-28}~e\,\mathrm{cm}$.  

% Distribution of weighted EDM values $d/\sigma_d$? $E^2$ plot of extracted frequencies? (Rosenberry did this).
%\begin{figure}[tb] %8.6cm
%\centering
%\includegraphics[width=\columnwidth]{histogram_bw}
%\vskip -0.14 truein
%\caption{\label{histogram} Histogram of the normalized residuals for all 120 EDM measurements, where $r_i=(d_i-d_A(\Xe))/\sigma_{d_i}$ for the weighted average of $d_A(\Xe)$ (Eq.~\ref{result}), $d_i$ is a single four-segment EDM measurement, and $\sigma_{d_i}$ is its uncertainty; the solid curve is a Gaussian with unit variance and area 120.}
%\vskip -0.14 truein
%\end{figure}
Further improvement to the polarization, SQUID dewar noise, measurement time, and increased electric field should result in an order of magnitude or more in $\Xe$ EDM sensitivity. The comagnetometer drift can be reduced with a more precise $\pi/2$ flip, tuning the ratio of $\Xe$/$\He$ polarizations, which was shown to be effective at the end of the 2018 campaign, and an optimized EDM cell shape~\cite{Limes2018}. Precise cell motion measurements are also essential.

This improved limit improves constraints on the low-energy CPV parameters developed in Refs.~\cite{Chupp2015,Chupp2017}, in particular lowering the limits on $\bar g_\pi^{0,1}$ and $\bar\theta$ by factors of two and $C_T$ by a factor of about five~\cite{Chupp2019inprep}; it can also be used to constrain the QCD axion contribution to EDMs by a factor of about five compared to that reported in~\cite{DzubaFlambaum2018}.
 
%While any further increase in sensitivity and upper limit will impact the global interpretation of EDM results \cite{Chupp2015}, an order of magnitude improvement would represent a significant advancement in our sensitivity to BSM physics. 
%This work also presents significant advances in comagnetometer analysis \cite{HeXeSupplement} and may have impact on other EDM and BSM searches including planned neutron EDM experiments with comagnetometers.
\begin{acknowledgments}
We wish to thank Patrick Pistel and Roy Wentz for excellence and innovation in glass blowing and cell construction. This work was supported in part  by NSF grant PHY-1506021, DOE grant DE-FG0204ER41331, Michigan State University, by  Deutsche Forschungsgemeinschaft grants TR408/12 and FA1456/1-1 and The Cluster of Excellence ``Origin and Structure of the Universe." WT acknowledges the support of a Humboldt Stiftung Fellowship.
\end{acknowledgments}

\end{document}